\documentclass[twocolumn,showpacs,preprintnumbers,amsmath,amssymb,prx]{revtex4-1}

\usepackage{mathrsfs}
\usepackage{graphicx}
\usepackage{dcolumn}
\usepackage{bm}
\usepackage{amsmath}
\usepackage{amsfonts}
\begin{document}

\title{Experimental Observation of Incoherent-Coherent Crossover and Orbital Dependent Band Renormalization in Iron Chalcogenide Superconductors}

\author{Z. K. Liu$^{1,2}$, M. Yi$^{1,2}$, Y. Zhang$^{1,3}$, J. Hu$^4$, R. Yu$^{5,6}$, J.-X. Zhu$^7$, R.-H. He$^8$, Y. L. Chen$^9$, M. Hashimoto$^{10}$, R. G. Moore$^1$, S.-K. Mo$^3$, Z. Hussain$^3$, Q. Si$^6$, Z. Q. Mao$^4$, D. H. Lu$^{10,\ast}$, Z.-X. Shen$^{1,2,\ast}$}
\affiliation{
$^1$ Stanford Institute for Materials and Energy Sciences, SLAC National Accelerator Laboratory, 2575 Sand Hill Road, Menlo Park, CA 94025, USA\\
$^2$ Geballe Laboratory for Advanced Materials, Departments of Physics and Applied Physics, Stanford University, Stanford, California 94305, USA\\
$^3$ Advanced Light Source, Lawrence Berkeley National Lab, Berkeley, California 94720, USA\\
$^4$ Department of Physics and Engineering Physics, Tulane University, New Orleans, Louisiana 70118, USA\\
$^5$ Department of Physics, Renmin University of China, Beijing 100872, China\\
$^6$ Department of Physics and Astronomy, Rice University, Houston, Texas 77005, USA\\
$^7$ Theoretical Division, Los Alamos National Laboratory, Los Alamos, New Mexico 87545, USA\\
$^8$ Department of Physics, Boston College, Higgins Hall 330N. Chestnut Hill, Massachusetts 02467, USA\\
$^9$ Physics Department, Clarendon Laboratory, University of Oxford, Parks Road, OX1 3PU, UK\\
$^{10}$ Stanford Synchrotron Radiation Lightsource, SLAC National Accelerator Laboratory, 2575 Sand Hill Road, Menlo Park, California 94025, USA
}

\begin{abstract}
  The level of electronic correlation has been one of the key questions in understanding the nature of superconductivity. Among the iron-based superconductors, the iron chalcogenide family exhibits the strongest electron correlations. To gauge the correlation strength, we performed systematic angle-resolved photoemission spectroscopy study on the iron chalcogenide series Fe$_{1+y}$Se$_x$Te$_{1-x}$ (0$<$x$<$0.59), a model system with the simplest structure. Our measurement reveals an incoherent to coherent crossover in the electronic structure as the selenium ratio increases and the system evolves from the weakly localized to more itinerant state. Furthermore, we found that the effective mass of bands dominated by the d$_{xy}$ orbital character significantly decreases with increasing selenium ratio, as compared to the d$_{xz}$/d$_{yz}$ orbital-dominated bands. The orbital dependent change in the correlation level agrees with theoretical calculations on the band structure renormalization, and may help to understand the onset of superconductivity in Fe$_{1+y}$Se$_x$Te$_{1-x}$.
\end{abstract}

\date{\today}

\maketitle

\section{Introduction}
The nature and role of many-body interaction has been a crucial yet unsettled question in the recently discovered iron-based superconductivity \cite{Pag2010,johnston2010,yin2011}. Among all the iron-based superconductors, the correlation level in the iron chalcogenide Fe$_{1+y}$Se$_x$Te$_{1-x}$ (11 system) has been predicted to be one of the strongest \cite{yin2011,aichhorn2010,yin2012}, which is confirmed by transport \cite{liutj2010,nojit2012}, neutron scattering \cite{lishiliang2009}, optical spectroscopy \cite{chengf2009}, and photoemission spectroscopy \cite{tamai2010,yamasaki2010,maletz2014} experiments. The mechanism for strong correlation in the parent compound of the 11 system, Fe$_{1.02}$Te was addressed in our previous angle-resolved photoemission spectroscopy (ARPES) study: the electronic structure in the antiferromagnetic (AFM) phase is featured by the characteristic ``peak-dip-hump'' features and quasiparticle dispersion with huge band renormalization ($\sim$90), which we attribute to coherent polarons formed by the interplay of large magnetic moment and electron-phonon coupling \cite{liuzk2013}. The coherent polaronic behavior naturally explains the metallicity in the AFM state of Fe$_{1.02}$Te. However, there is up to date no systematic study on the evolution of the correlation strength with the change of selenium ratio from the metallic AFM phase (x$<$0.1) to the weakly localized phase (0.1$<$x$<$0.28) and finally the superconducting/metallic phase (x$>$0.28) \cite{liutj2010} (Fig. 1(a)), where the correlation level is described by large band renormalization, reported to be 6$\sim$20 for the optimally doped FeSe$_{0.45}$Te$_{0.55}$ (Tc$=$14.5 K) \cite{tamai2010}.

In this work, we present a systematic study of the electron correlation effect using ARPES on a series of Fe$_{1+y}$Se$_{x}$Te$_{1-x}$ samples with increasing selenium ratios (y$<$0.02, x$=$0, 0.11, 0.2, 0.25, 0.28, 0.35, 0.44, 0.59). Our results show that the electronic structure of x$=$0.11 sample in the weakly localized phase is similar to that of Fe$_{1.02}$Te (x$=$0 sample) above the AFM transition temperature (T$_N$). With higher selenium ratio, the spectral weight of the coherent quasiparticles becomes increasingly pronounced, indicating an incoherent to coherent crossover in the electronic structure. Furthermore, we find that the effective mass renormalization of the bands dominated by the d$_{xy}$ orbital character decrease with selenium substitution, while those with the d$_{xz}$/d$_{yz}$ character do not show much change. Our results reveal an orbital dependent decrease of electronic correlations as superconductivity emerges in the iron chalcogenide Fe$_{1+y}$Se$_{x}$Te$_{1-x}$. Such evolution of orbital dependent electronic correlation effect is observed so far only in iron chalcogenides, making them a unique family to study the interplay between strong correlations, multi-orbital physics and superconductivity in iron-based superconductors.

\begin{figure*}[t]
\center
\includegraphics[width=0.6\textwidth]{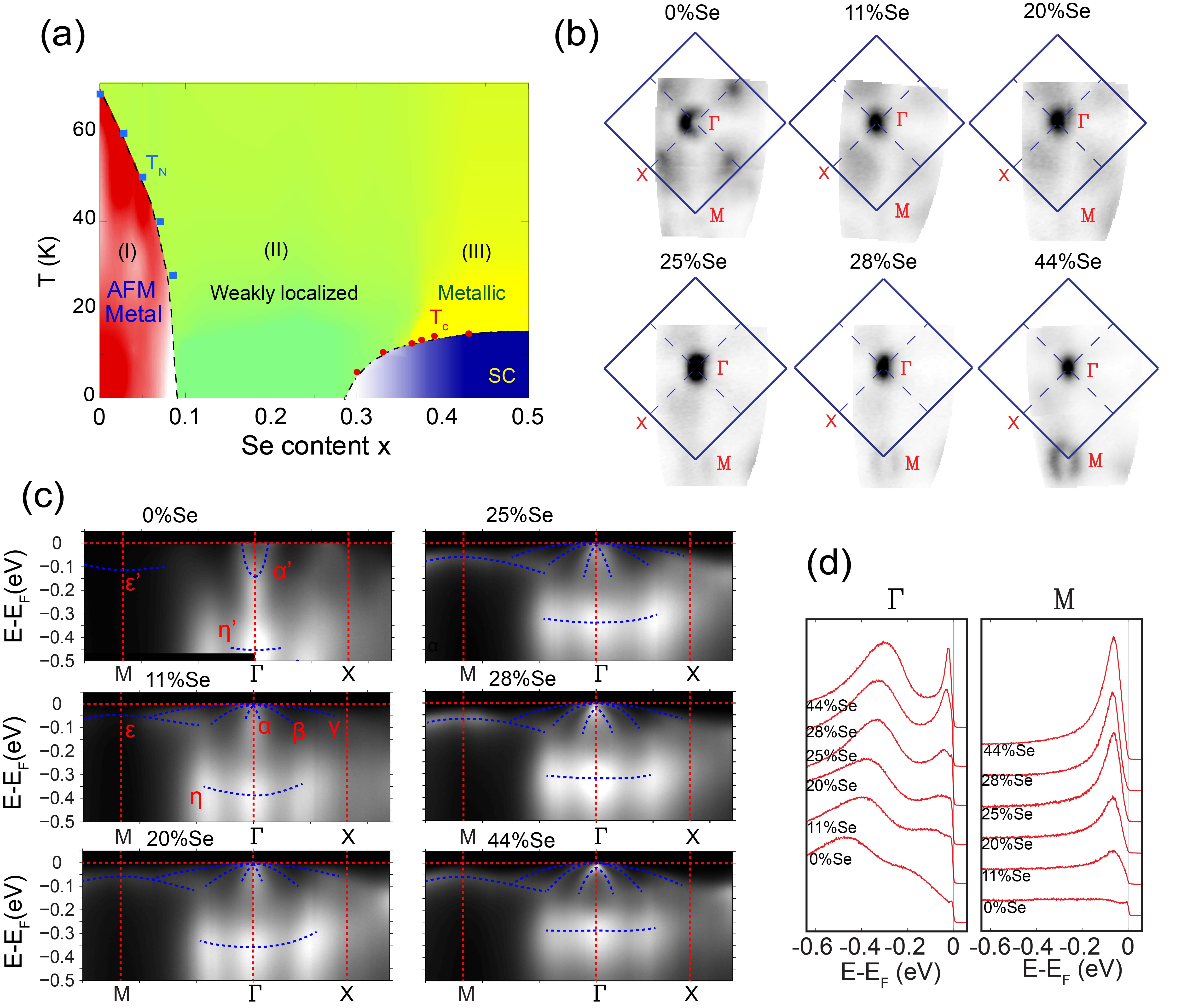}
  \caption{(a) Phase diagram of Fe$_{1+y}$Se$_x$Te$_{1-x}$ adapted from \cite{huj2013}. T$_N$ and T$_c$ represent the Neel temperature and onset superconducting transition temperature probed by specific heat, neutron scattering and magnetic susceptibility measurements. (b) Fermi surface map of Fe$_{1+y}$Se$_x$Te$_{1-x}$ with various x measured with 22 eV excitation energy at T$=$10 K. The photoemission intensity is integrated within a 20 meV window around the Fermi level. (c) Photoemission intensity of the cut along the M-$\Gamma$-X direction. Dotted curves are eye guides of different band dispersions. (d)  EDCs at both the $\Gamma$ and M points from Fe$_{1+y}$Se$_x$Te$_{1-x}$ with different x.}

\end{figure*}

\section{Experimental Results}
High quality Fe$_{1+y}$Se$_x$Te$_{1-x}$ single crystals were synthesized using flux method \cite{liutj2009}. Excess Fe ratio was kept as low as possible and was determined by energy-dispersive X-ray spectrometry to be around 2$\%$. ARPES measurements were performed at beamline 5-4 at Stanford Synchrotron Radiation Lightsource (photon energy h$\nu$=22$\sim$26 eV) and beamline 10.0.1 at Advanced Light Source, LBNL (photon energy h$\nu$=50 eV). The samples were cleaved \emph{in situ}, and measured in ultrahigh vacuum with a base pressure better than 3$\times$10$^{-11}$ torr, and data were recorded by a Scienta R4000 analyzer at 10 K sample temperature. The energy (angular) resolution was 8 meV (0.2$^{\circ}$, i.e., $\sim$0.008 {\AA}$^{-1}$ for photoelectrons generated by 22$\sim$26 eV photons) for the SSRL setup, and 15 meV (0.2$^{\circ}$, i.e., 0.012 {\AA}$^{-1}$ for photoelectrons generated by 50 eV photons) for the ALS setup.

    \begin{figure*}[t]
    \center
    \includegraphics[width=0.6\textwidth]{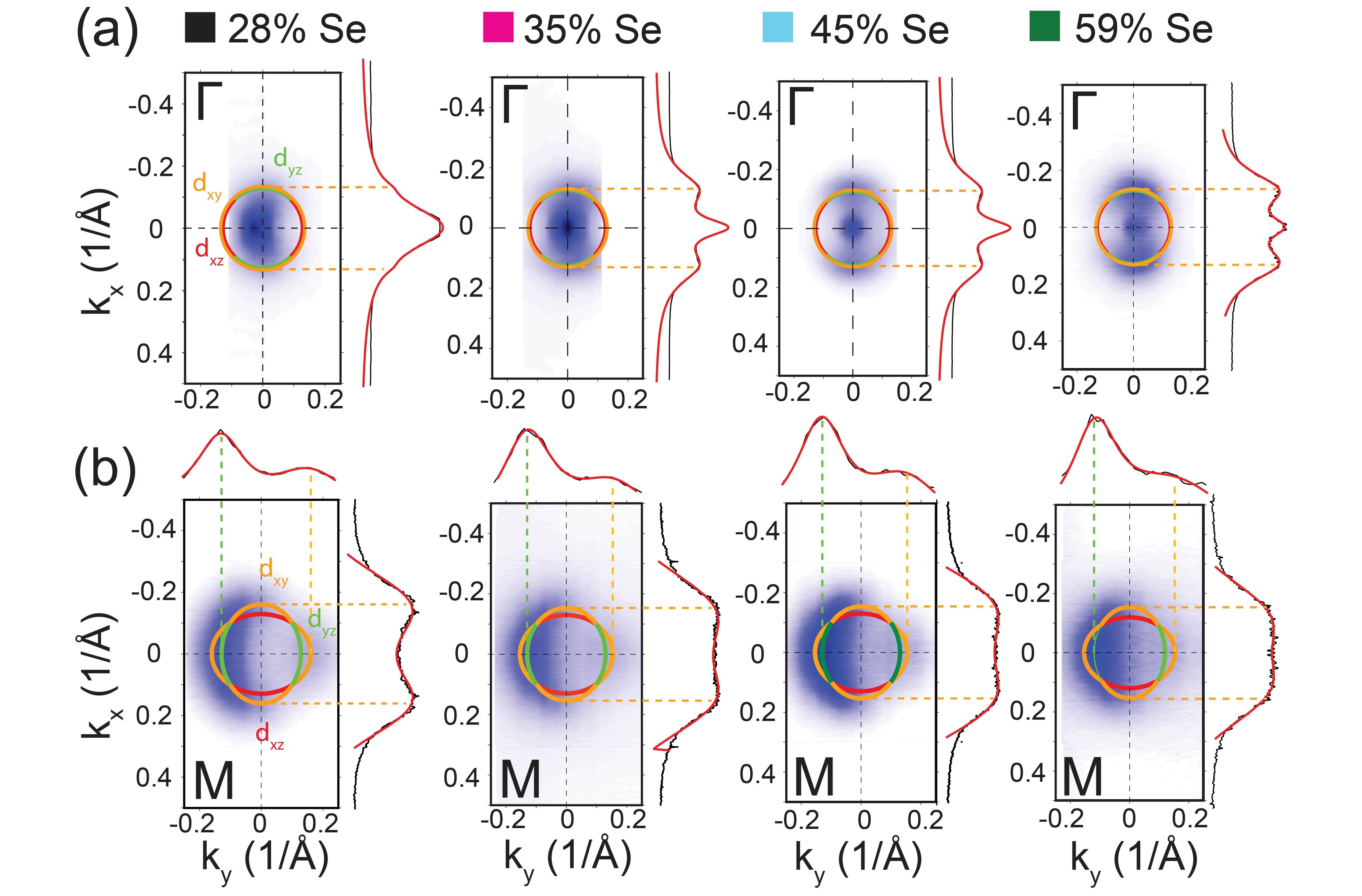}
     \caption{(a) Fermi surface maps of features around the $\Gamma$ point from Fe$_{1+y}$Se$_{x}$Te$_{1-x}$ with various x measured with 22 eV excitation energy at T=10 K. Schematic of the FSs is plotted on top of the map with different colors representing bands with different orbital characters. The momentum distribution curve (black curve) along k$_{y}$=0 is plotted on the right side of each figure, together with the three-Lorentzian fitting curve (red curve) showing the actual positions of the Fermi surface crossing (k$_{F}$). (b) Fermi surface map of features around the M point from Fe$_{1+y}$Se$_{x}$Te$_{1-x}$ with various x measured with 26 eV excitation energy at T=10 K. Schematic of the Fermi surface is plotted on top of the map with different colors representing bands with different orbital characters. The momentum distribution curve (black curve) along k$_y$(k$_x$)=0 is plotted on the right (top) side of each figure, together with the two-Lorentzian (Lorentzian+Gaussian) fitting curve (red curve) showing the actual positions of the Fermi surface crossing (k$_F$). }
    \end{figure*}

    \subsection{Incoherent-Coherent Crossover}
    The measured Fermi surfaces (FSs) and band dispersions along the high symmetry M-$\Gamma$-X directions (the high symmetry points are defined in the Brillouin zone in the reciprocal space of 2-Fe unit cell as shown in Fig. 1(b)) for samples with selenium ratios between 0 and 0.44 are plotted in Fig. 1(b) and (c), respectively. Firstly, the electronic structure of x=0 and x=0.11 at low temperatures are drastically different. Fe$_{1.02}$Te in the AFM phase is characterized by an electron pocket at $\Gamma$, finite spectral intensity around X and almost no intensity at M (see \cite{liuzk2013} for details). At the selenium ratio of 0.11 where the system is in the weakly localized state, the electron-like pocket at $\Gamma$ becomes hole-like, producing a peanut-like shape on the FS. Bands at M start to be noticeable while the spectral weight around the X point weakens. Notably, such a doping evolution in the electronic structure is very similar to that of the temperature induced change in Fe$_{1.02}$Te from below to above the AFM transition temperature T$_N$ (see \cite{liuzk2013}), hence could be understood as the result of electronic band reconstruction across the AFM phase transition.

    When the selenium ratio increases from 0.11 to 0.44, the changes in the electronic band structure become more gradual. From the FS mapping (Fig. 1(b)), we see that the feature at M becomes stronger in intensity with a gradual emergence of electron-like pockets, while the intensity at X fades out. Such a FS evolution may be closely related to the suppression of the ($\pi$, 0) short range magnetic order and the enhancement of the ($\pi$, $\pi$) magnetic fluctuation with selenium substitution, as reported in the neutron scattering experiments in the weakly localized phase of Fe$_{1+y}$Se$_x$Te$_{1-x}$ \cite{liutj2010,xuzj2010}. At the same time, the band dispersions do not have drastic changes in energy, except for the electron-like band around $\sim$400 meV below E$_{F}$ at $\Gamma$ (labeled as the $\eta$ band in Fig. 1(c) with dominantly d$_{z^2}$ orbital character), which shifts systematically from -400 meV to -300 meV, as can be also seen in the energy distribution curves (EDCs) at $\Gamma$ plotted in Fig. 1(d). This band shift is well captured by the density functional calculation results \cite{subedia2008}.

    \begin{figure*}[t]
    \center
    \includegraphics[width=0.8\textwidth]{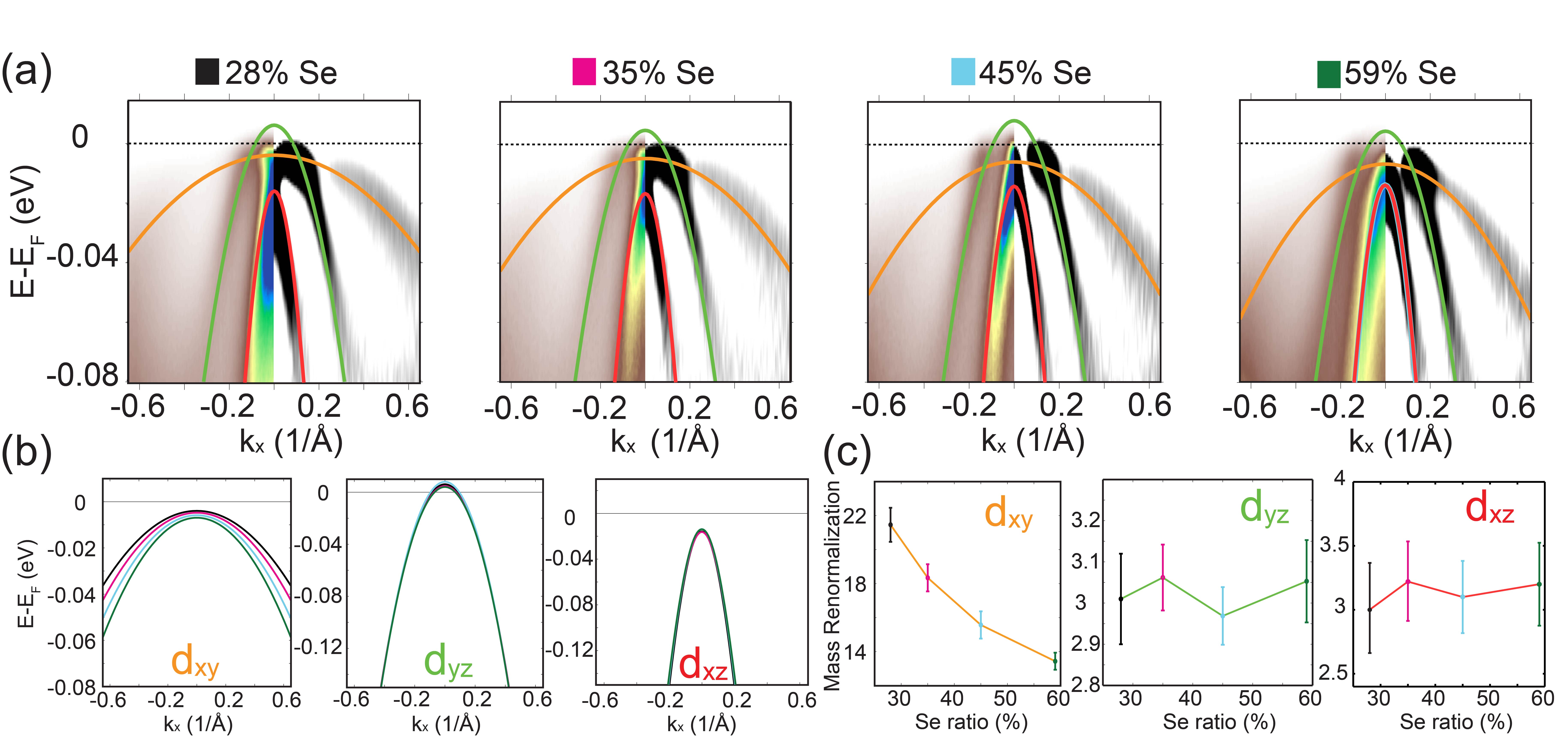}
      \caption{(a) Photoemission intensity of the cut around $\Gamma$ along the $\Gamma$-M direction for Fe$_{1+y}$Se$_x$Te$_{1-x}$ with various x. The original data is plotted on the left side of the each panel while the second derivative of the same data is plotted on the right side for better visualization of each band. The extracted dispersion for each band is plotted on top of each image plot with different colors representing different dominating orbital characters. (b) Stack plot of fitted band dispersions. Bands with different orbitals are plotted in different panels. In each panel different colors represent samples with different x. (c) Plot of the extracted effective band mass renormalization factor versus the selenium ratio. Band mass with different orbitals are plotted in different panels.}
    \end{figure*}

    More significant change occurs in the evolution of the spectral weight. As the selenium ratio increases, we found that in the high symmetry cuts (Fig. 1(c)) the broad and smeared dispersions become sharper, indicating that the spectral weight of the coherent quasiparticles becomes stronger over the incoherent background, as is clearly shown by the EDC evolution at both the $\Gamma$ and M points (Fig. 1(d)). The increase of the spectral weight of the coherent quasiparticles are very similar to ARPES observations of the doping-dependence of other strongly correlated materials (e.g., cuprates \cite{shenk2004,armitage2002}); it is a direct manifestation of the incoherent to coherent crossover behavior of the electrons, concomitant with Fe$_{1+y}$Se$_x$Te$_{1-x}$ evolving from the weakly localized phase to the metallic phase when the selenium ratio increases from 0.11 to 0.44. From 0.44 to 0.59, our measured quasiparticle spectral weights are fluctuating due to sample quality variations rather than showing the systematic trends of evolution.

    \begin{figure*}[t]
    \center
    \includegraphics[width=0.8\textwidth]{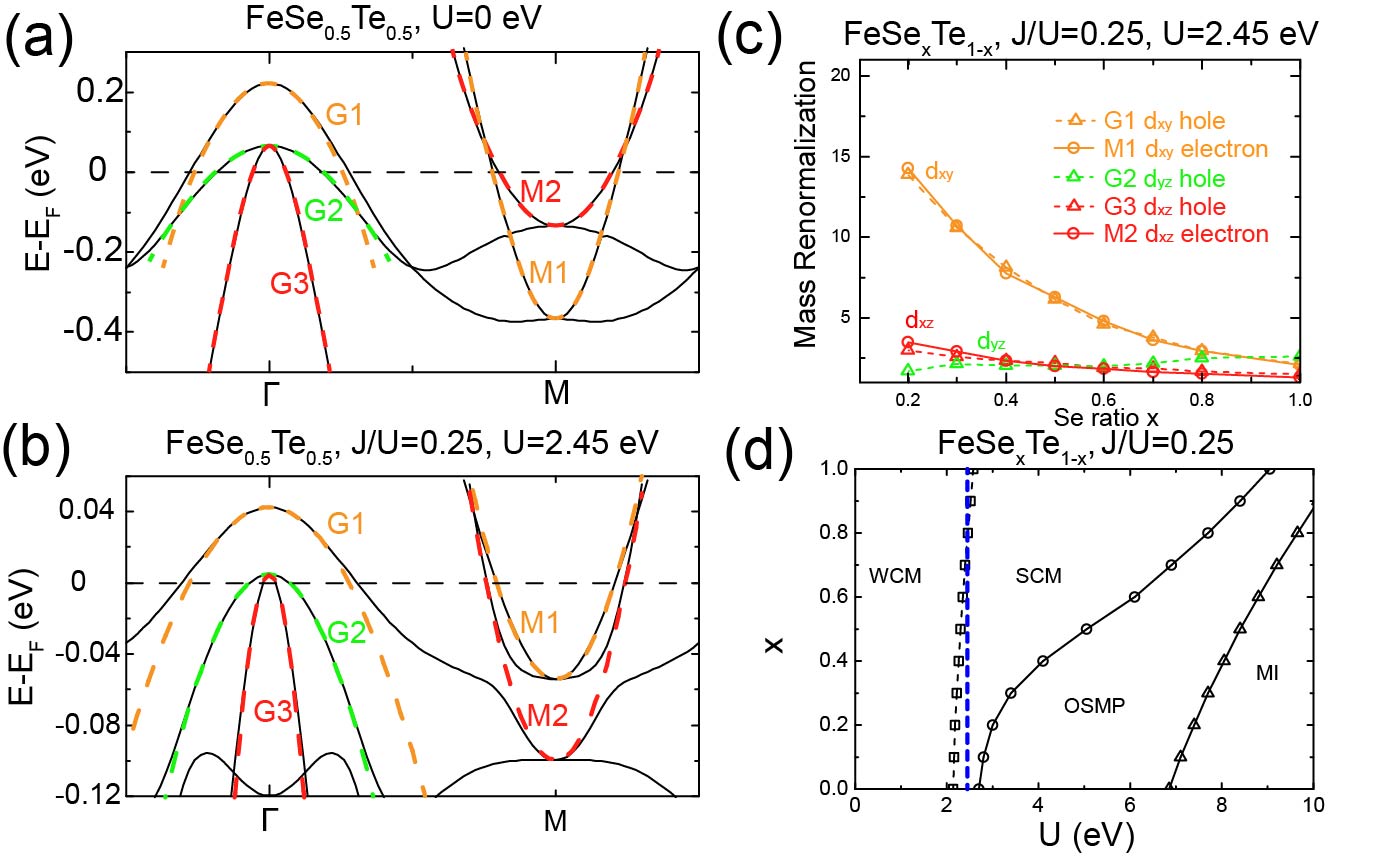}
      \caption{(a)-(b), Band structure calculation results on FeSe$_{0.5}$Te$_{0.5}$ with Coulomb repulsion U$=$0 and Hund's coupling J$=$0 for panel (a) and U$=$2.45eV, J/U$=$0.25 for panel (b). G1-G3 labels the three hole bands and M1-M2 labels the two electron bands. Dotted color curves indicate the extraction of the effective mass of different orbitals by parabolic curve fittings to portions of dispersion near the band top (bottom). Orange, green and red curves represent the d$_{xy}$, d$_{yz}$ and d$_{xz}$ orbitals. The breaking of dispersion near M is due to the band hybridization between the d$_{xz}$ electron band and d$_{xy}$ hole band. (c) Calculated mass renormalization for different bands from samples with different selenium ratio x in FeSe$_x$Te$_{1-x}$. The mass renormalization is calculated by dividing the effective mass from the calculated band structure with U$=$2.45eV, J/U$=$0.25 by that from U$=$J$=$0. Different marks with color indicate the renormalization of different bands, and different colors label different orbitals. (d) Calculated phase diagram of FeSe$_x$Te$_{1-x}$ as a function of selenium ratio x and Coulomb interaction U. The blue dotted line denotes where U$=$2.45eV and the calculated mass renormalization best agrees with the actual experimental results. WCM: weakly correlated metal. SCM: strongly correlated metal. OSMP: orbital selective Mott phase. MI: Mott insulator.}

    \end{figure*}

    \subsection{Orbital Dependent Band Renormalization}
    Such increase of coherence could not be interpreted as the effect of impurity scattering since the level of disorder actually increases with selenium substitution in Fe$_{1+y}$Te and maximizes at Fe$_{1+y}$Se$_{0.5}$Te$_{0.5}$ \cite{rajsv2011}. Rather, the crossover behavior strongly suggests the change in the electronic correlation strength. To address this problem, we performed detailed analysis on the electronic structure evolution of samples with selenium ratios from x$=$0.28 to 0.59 (the spectral weight of the dispersions for x$<$0.28 is too small, making such analysis difficult). Fig. 2 plots the measured FSs at both the $\Gamma$ (a) and M (b) points. At the $\Gamma$ point, we find the ``peanut shape'' in the FS is part of a circular hole pocket with mixed orbital content of d$_{xy}$ and d$_{xz}$/d$_{yz}$ due to band hybridization (see supplemental material, part II).  The reason we did not observe a full circle is due to the suppression from the matrix element (see supplemental material, part I). The intensity at exact $\Gamma$ comes from the band top of the inner d$_{xz}$/d$_{yz}$ band. At the M point, we observe two intersecting ellipses, with different segments coming from different orbitals. Some parts of the two ellipses are not visible in the data due to the suppression from the matrix element. To determine the Fermi pocket size, we find the maximum and minimum openings of the pockets from fitting the contour plots. Schematics of the FSs can then be drawn by considering the four-fold crystal symmetry of the tetragonal state. The carrier concentration level can then be evaluated by counting the Fermi surface volume. The calculated net doping level is 0.003$\sim$0.012 electron/Fe, indicating that the 11 system is almost electron-hole balanced with the additional electrons may come from the excessive interstitial Fe atoms \cite{liutj2009}. Furthermore, we do not observe any obvious change in either the electron or hole pockets for samples with different selenium ratios, confirming the isovalent nature of the selenium/tellurium substitution.

    The electronic band dispersions near the Fermi level do not shift noticeably in position but exhibit systematic change in curvature. We extracted the dispersions by locating peak positions from both the energy and momentum distribution curves. Around the $\Gamma$ point we could identify three different hole bands which are dominated by different orbital characters (see supplemental material, part I). By fitting each of the dispersion to a parabolic curve, we can extract the effective mass of each band (Fig. 3 (a), (b)). As Fig. 3 (c) shows, the d$_{xy}$ band has the largest effective mass renormalization compared with density functional calculation \cite{subedia2008}, while the other bands have much smaller renormalization factors ($\sim$3 for the d$_{yz}$ and the d$_{xz}$ band). Moreover, only the renormalization factor of the d$_{xy}$ band shows significant doping dependence, decreasing from 22 to 14 as the selenium ratio increases from 0.28 to 0.59. For the d$_{xz}$/d$_{yz}$ bands, the extracted effective masses do not change noticeably over this doping range.

    We applied the same analysis to the electron pockets at M and observed similar behavior (see supplemental material, part III). All together, we have found that bands dominated by the d$_{xy}$ orbital character have a much larger renormalization comparing to bands of other orbitals, and the renormalization monotonically reduces as the selenium ratio increases while the other orbitals do not have such effect. Therefore, by using effective band mass as the correlation level indicator, we have discovered there is a reduction of (yet still strong) correlation as selenium replaces tellurium in Fe$_{1+y}$Se$_{x}$Te$_{1-x}$, and this correlation reduction has most significant effect in bands with the d$_{xy}$ orbital character.

\section{Theoretical Interpretation}

    Comparing to the band renormalization change that happens to all the t$_{2g}$ bands in Co-doped BaFe$_{2}$As$_{2}$ system \cite{sudayama2011}, such orbital dependent renormalization evolution is a unique feature for the 11 system. So far, there have been several theoretical works discussing the level of correlation and unique orbital dependent physics in the 11 system. In a theoretical study based on dynamical mean-field theory, ref \cite{yin2011} has pointed out that the correlation effects in Fe pnictides/chalcogenides come from the Hund's coupling and that the structural parameters have a strong impact on the overall correlation strength and orbital-selectivity. The longer Fe-chalcogen bond length compared to that of Fe-pnictigen would result in more localized electrons. Meanwhile, the Ch-Fe-Ch bond angle, which controls the crystal field splitting, is much smaller from that of an ideal tetrahedron in iron chalcogenides. As a result, electrons in the in-plane d$_{xy}$ orbitals are more localized in iron chalcogenides compared to other orbitals. Given the structural sensitivity, the substitution of smaller selenium atoms for bigger tellurium atoms would modify the structural parameters and decrease the correlation level and orbital selectivity. In another work based on a slave-spin mean-field method \cite{yurong2013,yurong2011}, the authors proposed that due to the strong intra-orbital Coulomb repulsion U and Hund's coupling J, the iron chalcogenide family is in proximity to an orbital selective Mott phase (OSMP), where the d$_{xy}$ orbital is Mott-localized while the other orbitals remain itinerant.

    Both works would explain our observations on the correlation level change and orbital selectivity in FeSe$_{x}$Te$_{1-x}$. As an illustration, we applied the model from \cite{yurong2013,yurong2011} to the FeSe$_{x}$Te$_{1-x}$ family and the calculation results are summarized in Fig. 4. With the inclusion of moderate U and J (U$=$2.45 eV, J/U$=$0.25) in the model, the dispersion of FeSe$_{0.5}$Te$_{0.5}$ [Fig. 4(b)] is found to be greatly renormalized comparing to the U$=$J$=$0 case [Fig. 4(a)]. The calculated mass renormalizations at the same value of U and J for different x values in FeSe$_{x}$Te$_{1-x}$ well reproduce the change in the overall correlation strength and orbital-selectivity as observed [Fig. 4(c)]. The agreement with the experimental result shows that FeSe$_{x}$Te$_{1-x}$ are overall in the strongly correlated metal phase and loses correlation with increasing x [Fig. 4(d)]. The calculation further proposed that the correlated metal phase is in proximity to an orbital selective Mott phase and raising temperature is one potential path to enter such phase \cite{yiming2014}.

\section{Discussions}
    It should be noted that our observation of the correlation evolution of Fe$_{1+y}$Se$_{x}$Te$_{1-x}$ has only extended to selenium ratio up to 0.59. Single crystals with selenium ratio higher than that have been found to be hard to stabilize \cite{mizuy2010}. However, in our recent ARPES measurement, we have observed very renormalized d$_{xy}$ hole bands with renormalization factor $\sim$10 for K$_{x}$Fe$_{2-y}$Se$_{2}$ \cite{yiming2014,yiming2013}. In addition, a recent ARPES report on single crystal of FeSe found the d$_{xy}$, d$_{yz}$ and d$_{xz}$ hole band renormalizations to be 9, 3 and 3.7 respectively, fully consistent with our observations of the trend \cite{maletz2014}. Therefore, the large d$_{xy}$ orbital band renormalization appears to be universal to all iron chalcogenides, making it unique among all iron-based superconductors \cite{yiming2014}.

     The nature of the strong correlation may be critical to the understanding of the superconductivity in iron chalcogenides. In Fe$_{1+y}$Se$_{x}$Te$_{1-x}$, the level of correlation seems to be the primary tuning factor for superconductivity since the doping level and the underlying Fermi surface topology do not change. For K$_{x}$Fe$_{2-y}$Se$_{2}$, where Tc is comparable to iron pnictides, the lack of hole pockets makes a weak-coupling Fermi surface nesting picture unlikely. Hence, the superconducting pairing mechanism may stem from strong correlations that lead to strong local pairing. As the d$_{xy}$ band is most sensitive to the change of correlation among all Fe 3d bands, its band renormalization would serve as an accurate gauge for the correlation level and pairing strength.

\begin{acknowledgments}
ARPES experiments were performed at the Stanford Synchrotron Radiation Lightsource and the Advanced Light Source, which are both operated by the Office of Basic Energy Sciences, U.S. Department of Energy. The Stanford work is supported by the US DOE, Office of Basic Energy Science, Division of Materials Science and Engineering, under award number DE-AC02-76SF00515. The work at Tulane is supported by the NSF under grant DMR-1205469 and the LA-SiGMA program under award $\sharp$EPS-1003897. The work at Rice has been supported by NSF Grant DMR-1309531 and the Robert A. Welch Foundation Grant No. C-1411. The work at Renmin University has been supported by the National Science Foundation of China Grant number 11374361, and the Fundamental Research Funds for the Central Universities and the Research Funds of Renmin University of China.
\end{acknowledgments}

\bibliography{FST_phasediagram}
\newpage

\end{document}